\documentstyle[12pt]{article}
\begin{document}

\begin{center}
{\Large {\bf Gap anisotropy in the angle-resolved photoemission spectroscopy of 
Bi$_2$Sr$_2$CaCu$_2$O$_{8+\delta}$ }}\\
\vspace{1.0cm} 
{{\bf Tulika Maitra$^{*}$}\footnote{email: tulika@phy.iitkgp.ernet.in} 
and {\bf A. Taraphder}$^{*\dagger}$\footnote{email: arghya@phy.iitkgp.ernet.in, 
arghya@cts.iitkgp.ernet.in}}\\ 

\noindent $^{*}$Department of Physics \& Meteorology and  
$^{\dagger}$Centre for Theoretical Studies, \\
Indian Institute of Technology,
Kharagpur 721302 India \\   

\end{center}  
\begin{abstract}

The gap anisotropy in Bi$_2$Sr$_2$CaCu$_2$O$_{8+\delta}$ is revisited 
in the framework of a d-wave scenario in view of the recent angle-resolved 
photoemission experiment. Based on a tight-binding fit to the 
normal state dispersion, a detail analysis on the effects of the inclusion of 
the next harmonic in the d-wave has been presented. Significant effect has 
been observed in the superconducting T$_c$. The density of states is linear 
at the nodes with enhanced weight, caused by a marked increase in the 
low energy excitaions which affect the thermodynamics considerably. The slope 
of the $\rho_s - T$ curve in the low temperature regime increases and 
the specific heat reflects the enhanced entropy at low temperatures. The 
leading edge of the ARPES energy distribution curves have been calculated
and found to shift towards higher energy. The effect of scattering by 
non-magnetic impurities in this context are also outlined. 

\end{abstract}  
\vspace{.5cm} 

\noindent {\bf Introduction}  
\vspace{.5cm} 

The nature of the superconducting gap anisotropy in the high temperature 
superconductors has attracted a lot of attention in the last few 
years\cite{van,leggett}.
A considerable progress has been made in the recent past with the resolution
and clarity of the angle-resolved photoemission spectroscopy (ARPES) having 
reached a very high level backed by careful and thorough analysis of the 
data\cite{nor0}. 
A general consensus seems to have emerged that the symmetry of the order 
parameter (OP) in most of the high temperature superconductors is predominantly
d-wave. In a very recent ARPES experiment on the underdoped Bi2212 system Mesot 
et. al.\cite{mesot} have reported an observable departure from a simple 
interpretation in terms of the usual d-wave symmetry. Inclusion of the next 
higher harmonic in the superconducting pairing function seems to give a better 
fit to their data. We develop the idea in order to look for other observable 
effects of such a term, and calculate its effects on various physical quantities
in the framework of the usual phenomenological theory\cite{nor1} in the weak 
coupling limit and suggest further experiments. We have calculated the changes 
in  superconducting transition temperature, nodal structure on the fermi 
surface (FS), the density of states (DOS) in the superconducting state and 
its effect on the specific heat. The energy distribution curves (EDC) (that 
are seen  in the ARPES experiments) are also obtained. 
Fluctuations are included in one loop level and the superfluid density 
(and hence the penetration depth) is obtained as a function of temperature. 
We observe the differences in slopes of $\rho_s (T)$ in the low temperature 
regime with increasing level of mixing of the higher harmonic. We also 
calculate the effect of both doping and the mixing of higher order terms 
on the specific heat. 
\vspace{.5cm} 

\noindent {\bf Model and Calculations} 
\vspace{.5cm} 

In the experiments of Mesot et al, ARPES results have been fitted 
to a d-wave gap function for several Bi2212 systems. The maximum gap 
values have been adjusted at each of the doping levels for the best fit, and  
the EDCs at different angles and the slope of the gap function close to the 
node have been carefully fitted. Clear deviation from the usual 
$d_{x^{2}-y^{2}}$ behaviour is observed and attempts were made to fit the 
data with an admixture of about 4-10\% of the next harmonic cos(6$\phi$). A 
much improved fit was indeed obtained with the inclusion of this higher order 
term. As one underdopes the system, the normal state resistivity 
is known to increase and effective screening becomes weaker. Under such a 
situation it is reasonable to assume that the higher harmonics in
the effective interactions have to be brought in\cite{scal}, and highly 
sensitive experiments would reveal the effects due to the growing range of 
interaction. In the spin fluctuation models too the pairing interaction
grows sharply in the momentum space with underdoping and long range effects 
in real space become increasingly relevant. 

We undertake to examine the ramifications arising out of the addition of
the higher harmonic with a model band structure that reproduces the observed 
FS and the location of the van-Hove singularity in Bi2212
accurately\cite{cgt,nor1}. The six parameters describing the hopping on 
the BiO plane used\cite{cgt,nor1} to obtain the fit were [0.131, -0.149, 0.041,
-0.013, -0.014, 0.013] (in eV), corresponding to a doping of $\delta=0.17$. 
The van-Hove singularity, due to the saddle points $\bf k=(\pi,0), (0,\pi)$ 
is 30meV below the FS, as observed in ARPES.   

Superconductivity occurs through attractive long range interactions of
which we keep only the near-neighbour part\cite{cgt,nor1}. Combined with the
fact that there is a reasonably strong on-site repulsion between the electrons
in the cuprates, forcing the pairing function $\Psi(r=0)=0$, such an 
interaction is known to support d-wave pairing\cite{scal,moat,kot}.
The vanishing of the on-site part of the pairing function implies that
the $\sum_{q} \Delta(\bf q)=0$, where the $\bf q$-sum runs over the entire 
Brillouin zone (BZ), thereby contributing equal regions of positive and
negative sign to the OP. The usual $d_{x^{2}-y^{2}}$ (or $cos(2\phi)$) OP 
admits of such constraints. All the higher harmonics, $cos[(2+4n)\phi]$ 
are also admissible under the symmetry restrictions of the d-wave. In the
underdoped systems, it is likely that higher neighbour interactions contribute
increasingly, but in the present analysis we kept only the first of such 
terms. We also include only the singlet component of the order parameter
in our analysis, as there is no reason so far to include the triplet part
in the parameter regions that one works with in these system\cite{nor1,scal}.  

The superconducting gap equation that is numerically solved is the usual 
mean-field factorized gap equation
$$\Delta_{\bf k}=\frac{1}{N}\sum_{\bf k'}V({\bf k-k'})
\frac{\Delta_{\bf k'}
tanh(\beta E_{\bf k'}/2)}{2 E_{\bf k'}},$$ 
\noindent where $V({\bf k-k'})$ is written in the separable form\cite{kot} 
$$V({\bf k-k'})=g\eta({\bf k}) \eta({\bf k'}).$$ 
Here we have chosen the basis function $\eta({\bf k})$ 
to be the $B_1$ representation of the one dimensional irreducible 
representations of $C_{4v}$, $\eta({\bf k})=\frac{1}{2}(cosk_{x}-cosk_{y})$ 
(the usual $d_{x^{2}-y^{2}}$ symmetry);  $g$ measures the strength of 
the attractive 
interaction. The triplet channels of pairing have not been considered 
in the foregoing\cite{leggett}. The value of $g$ was chosen (320 meV) 
such that the transition temperature $T_c$ at $\delta=0.17$ remains 
close to 80K (without any admixture of $cos(6\phi)$ term). In the 
spirit of Mesot et al., we introduced the next order term 
$cos(6\phi)$ in $\eta({\bf k})$ as $\eta({\bf k})=\alpha cos(2\phi)+
(1-\alpha)cos(6\phi)$, where $\alpha$ measures the relative contributions 
of the two terms.

The superconducting gap has been calculated for each filling ($\delta$) at 
different levels of mixing of the higher harmonic. As representative curves, 
we show in Fig.1 the ones with 0, 4, 10 and 20\% of mixing for $\delta=0.17$. 
The solutions of the gap function for different $\alpha$ show typical 
square root behaviour as $T\rightarrow T_c$. The transition temperature 
(and $\Delta(0)$) reduces considerably as $\alpha$ deviates from one (the 
inset to Fig. 1). The gap function with the same set of values of $\alpha$ 
are drawn in the Y-quadrant (Fig. 2) for a demonstration of how they change 
with the mixing of $cos(6\phi)$. Similar shifts have been seen by Mesot et al. 
as well.  

Calculation of the density of states (DOS) in the superconducting state 
with and without the higher harmonic term is straightforward. For $\alpha=1$,
one gets the usual d-wave DOS with $\rho(E) \propto |E|$ for $E\rightarrow 0$.
Addition of the higher harmonic term makes the rise steeper as there are
now more excitations available at lower energy though the low energy behaviour
of the DOS remains the same (linear). In Fig. 3 the DOS is shown for the 
normal and the broken symmetry states (with only $\alpha=1$, as the 
$\alpha=0.8$ curve is indistinguishable in the scale of that figure). The 
typical d-wave V-shaped DOS is obtained with the shoulders at $\pm\Delta$ 
around $E=0$. The inset shows for a comparison, in an enhanced scale of 
energy, the effect of the addition of the higher harmonic term. The pile-up 
of states at lower energy is quite evident.    

In order for an analysis of the ARPES spectra we take the point of 
view\cite{ding} that the spectral function $A({\bf k},\omega)$, convoluted 
with the fermi function $f(\omega)$ and the resolution of the detector, 
gives the EDC. The ARPES intensity (without the detector resolution) is given 
by $I_{0}f(\omega)A({\bf k},\omega)$ where, $I_{0}$ is a prefactor weakly 
dependent on $T$ and $\omega$. It depends on the incident photon energy, 
momentum and the (electron-phonon) matrix element between the initial and 
final states. The momentum resolution of the detectors used in experiments 
considered is about one degree in the BZ and is assumed constant over a 
circular window of 1$^{o}$ radius. With the incident photon energy around 20eV,
the energy resolution of the detector is taken to be a Gaussian of standard 
deviation 7meV, a value consistent with the present day experimental 
resolutions\cite{ding}. 

\noindent With all these taken into account, the ARPES intensity is given
by 
$$ I({\bf k},\omega)=I_{0}\int_{\bf k'}\int_{\nu}\tilde{G}(\omega-\nu)f(\nu)
A({\bf k'},\nu).$$
 
\noindent Here $\tilde{G}(\omega)$ is the Gaussian energy resolution function 
discussed above and the ${\bf k'}$-integration is within a circular radius 
of 1$^{o}$. The spectral function is the usual mean-field one 
$$A({\bf k},\omega)=u_{\bf k}^{2}\delta(\omega-E_{\bf k})+v_{\bf k}^{2} 
\delta(\omega+E_{\bf k})$$ 
\noindent where, $E_{\bf k}=\sqrt{\epsilon_{\bf k}^{2}
+\Delta_{\bf k}^{2}}$, and the coherence factors $u_{\bf k}^{2}=\frac{1}{2}
(1 +\epsilon_{\bf k}/E_{\bf k}), v_{\bf k}^{2}=\frac{1}{2}(1-\epsilon_{
\bf k}/E_{\bf k}).$
 
Taking the normal state dispersion $\epsilon_{\bf k}$ discussed above, 
$A({\bf k},\omega)$ at five representative angles (on the FS) $\phi=0,10, 23,
35, 45$ have been calculated and their frequency and momentum-resolved 
behaviour (Fig. 4) obtained. The angles are measured with respect to the 
line $(\pi,\pi) - (\pi,0)$ as is the practice\cite{nor1}. We have restricted 
ourselves to 
the Y-quadrant of the Brillouin zone in order to avoid the complications with   
the shadow bands that appear in the X-quadrant due to the presence of 
an incommensurate superlattice (along $\Gamma-Y$ direction) in the 
BiO planes\cite{mesot,nor1}. We choose a point on the FS (at the above 
angles), and perform the resolution averaging around that point in momentum 
and frequency. Extreme care is required in evaluating the averages on the
FS where the band is most dispersive, i.e., around 45$^o$. Panel 1 in 
Fig. 4 shows frequency averaged $A({\bf k}, \omega)$ 
and panel 2, the momentum-averaged one. The procedure has been repeated 
for $\alpha=1.0$ and 0.90 to show the effects of the higher harmonic 
on the gap function as one moves along the FS. 

With the momentum and frequency averaged $A({\bf k}, \omega)$ obtained, it is
easy to calculate the ARPES intensity around different points on the FS. 
The curves at different angles are shown in panel 3 of Fig. 4. The maximum 
gap is taken to be 30 meV, the same as used by Mesot et al, for demonstration.
The temperature used is 10K, a typical value used in experiments at low 
temperatures. 

It easily follows from the above discussions that the thermodynamic
properties will be affected by the addition of the higher harmonics. 
We calculate the specific heat as a function of temperature (Fig. 5) in
the superconducting state by taking a derivative of the entropy with
temperature. Figs. 5 (a), (b) and (c) show the specific heat for different
values of $\alpha$ at several levels of doping. It is clear from the 
graphs (a) - (c) that in order to account for the extra entropy, the 
curves move up as higher harmonics are brought in and then the area
is conserved with reduced transition temperatures. As the doping increases, 
the curvatures of the Fermi surface affect the specific heat as seen
in the normal state specific heat curve in (c). At T$_c$ the specific heat 
jumps to its normal state value as is typical of a second order transition.
 
The calculation of superfluid density $\rho_s$ has been performed using 
the standard techniques of many body theory\cite{cgt,scal1}. 
The diamagnetic and paramagnetic contributions to the current (and hence
the phase stiffness) are calculated in the linear response by first making
a Peierls substitution $t_{ij}=t_{ij}exp(ie/c\hbar)\int_{{\bf r}_{j}}^{
{\bf r}_{i}}{\bf A}.d{\bf l}$ in the hopping matrix element. 
The paramagnetic contribution at long wavelengths (via excitations above the 
condensate) in the linear response theory is obtained in terms of the 
correlation function 
$${\bf j}^{x}_{para}({\bf q})=-\frac{i}{c}\,\,lim_{q\rightarrow 0}
lim_{\omega\rightarrow 0}\int d\tau\theta(\tau)e^{i\omega\tau} \langle 
[j_{x}^{para}({\bf q},\tau),j_{x}^{para}(-{\bf q},0)]\rangle 
{\bf A_{x} ({\bf q})}.$$ 
The correlation
function on the RHS is calculated at the one-loop level. Calculation of the 
diamagnetic contribution (from the Meissner condensate) is straightforward: 
${\bf j}^{x}_{dia}({\bf q})= -\frac{e^{2}}{N\hbar^{2}c}\sum_{{\bf k},\sigma}
\langle c^{\dagger}_{{\bf k},\sigma} c_{{\bf k},\sigma}\rangle{\bf A_{x} 
({\bf q})}.$ 
\noindent In the resulting expression, the OP values were taken as their 
mean-field unrenormalized (by the fluctuations) ones, a procedure that is 
known to work\cite{cgt} except very close to T$_c$. The resulting $\rho_s-T$ 
curves are shown in Figs. 6  for $\alpha=1, 0.9$ and 0.8 at different doping
levels. Note that the calculations were done in the gauge $A_{y}=0$ and
the gauge invariance is restored if vertex corrections are included. Such an 
approximation entails neglecting the vortex-like fluctuations in the 2D 
model\cite{scal1}. The above expression for $\rho_{s}$ is derived for an 
isotropic order parameter but is expected to work quite well even in the
anisotropic case at hand as the asymptotic form of the vortex-vortex interaction
(at high vortex density) is logarithmic\cite{ren} and corrections to the 
expression above due to these fluctuations are indeed small.  
\vspace{.5cm} 

\noindent {\bf Results and Discussion} 
\vspace{.5cm} 

The nature of the OP gleaned from Fig. 1 shows a sensitive dependence on 
$\alpha$, going down as mixing increases. The cos($6\phi$) term changes sign 
four times now in each quadrant and such rapid changes average out to a smaller
value\cite{moat}. This is more pronounced if the next higher harmonic 
($cos(10\phi)$) is 
introduced\cite{tobe}. As $\alpha$ decreases, the $\Delta$ versus $\phi$ curve 
becomes flatter around the node (Fig.2) enhancing the quasiparticle excitations
above the condensate. Such excitations will reflect in a reduced T$_c$ (and 
$\Delta(0)$) and pile up of states in the DOS at low energies (Fig. 3). This 
will, of course, affect the thermodynamics considerably. For example, though
the specific heat follows the typical (mean-field) T$^{2}$ 
behaviour\cite{moler} of a d-wave superconductor at low temperature, the 
increased entropy at lower energy (for $\alpha$ deviating from one) manifests 
itself in the specific heat curves (Fig. 5). 

The ARPES line shapes are shown in Fig. 4. The frequency and momentum-
averaged spectral function is plotted in panel 1 and panel 2 at different 
angles relative to ($\pi,\pi$) on the FS. $A({\bf k}, \omega)$  (not shown)
has the usual delta function behaviour, two sharp peaks separated by 2$\Delta$
at $\phi=0$ and closing in as $\phi$ increases and finally merging at 
$\phi\rightarrow 45^{o}$. The frequency broadening (panel 1) is a mere 
consequence of the Gaussian averaging procedure on the $A({\bf k}_{F},
\omega)$. Strong angle dependence is observed in panel 2 where momentum 
averaged $A({\bf k}_{F},\omega)$ is plotted for different $\phi$. The OP does 
not change within the ${\bf k}$-window for small angles, while for large 
angles, the effects are very strong. At large angles the band is highly
dispersive and large changes in energy occur within the ${\bf k}-$window. 
The $\bf k$-dependence of the OP further enhances this angle
dependence since close to the node $\Delta_{\bf k}$ varies linearly with
${\bf k}$ while at the gap-maximum the variation is a weaker quadratic one.   
The $\bf k$-averaged spectral weight for $\alpha=0.9$ shows perceptibly 
smaller gap as $\phi$ increases. Such a reduction of the gap has been
observed by Mesot et al. The effect of $\alpha$ is clearly visible in 
the EDCs in panel 3. The leading edges for the curves corresponding 
to $\alpha=1$ and 0.90 move continuously away from the fermi energy as $\phi$ 
decreases, the curve for $\alpha=1$ moving more rapidly. This is the 
scenario depicted in Mesot et al. for seven different angles, where the best 
fit is obtained with $0.89 \le \alpha \le 0.96$ for different
samples. The peaks observed in our EDCs are due to the mean-field
pile-up of states, and not due to electronic correlations\cite{ran}.  

The superfluid density $\rho_s$ is proportional to $\lambda^{-2}$ ($\lambda$ 
is the penetration depth) and has a power law dependence on 
temperature\cite{hirs} at low temperatures. Figs. 6 (a)-(c) show $\rho_s$ 
for three representative $\alpha$ (1.0, 0.90 and 0.80) and is seen to fall 
off faster with the inclusion of the higher harmonics. As observed 
earlier\cite{hirs,hardy}, the curves are linear to a high degree close to 
zero temperature. We observe that the slope of $\rho_s (T\rightarrow 0)$  
decreases monotonically with increasing $\alpha$ (i.e., increasing T$_c$). 
As temperature rises, excitations from the condensate tend to decrease 
$\rho_s$ at the expense of normal quasiparticles above the condensate. 
The gradual flattening of the $\Delta-\phi$ curve with decreasing $\alpha$ 
around the node makes quasiparticle excitations more accessible at lower 
temperatures, causing a faster descent of $\rho_s$ with temperature. We have
also calculated $\rho_s$ at different doping levels and find
that the slope increases as one underdopes, consistent with the
observations of Mesot et al.  

It is known that non-magnetic impurities act as pair breaking scatterers
for a d-wave superconductor\cite{hirs1}. Repeated scattering even with
small momentum transfer on the fermi surface between lobes
of opposite sign in the Brillouin zone effectively reduces the average gap
value\cite{cmv} thereby reducing the T$_c.$ Owing to the presence of the
$cos(6\phi)$ term, such pair-breaking processes are clearly going to be more 
efficient and will reduce the transition temperature as the higher harmonic 
components increase\cite{tobe}. The pair breaking effect of momentum
dependent impurity potential and anisotropic gap including the higher 
harmonics have been studied recently\cite{haas} by solving the Abrikosov-Gorkov 
equations in the T-matrix representation where it has been found that states 
begin to appear in the gap as the impurity potential as well as the 
higher harmonic component is increased thereby reducing both the 
superconducting gap and the transition temperature. 
 
In conclusion, we have worked out the effects due to the increasing presence 
of higher harmonics observed recently in a d-wave superconductor on underdoping.
Based on the standard phenomenological theory for d-wave 
superconductors\cite{nor1,kot}, we observe that the transition temperature, 
density of states, specific heat, superfluid density, ARPES intensity and 
the slope of the OP at the node bear clear and detectable signature of this 
higher order term. Although the interaction between quasiparticles have not 
been included in the above, the conclusions drawn remain valid 
qualitatively on strong physical grounds as has been shown in a number of 
occasions earlier\cite{nor1,cgt}. The predictions made here are easily 
verifiable experimentally and will shed light on the nature and strength
of the higher order term claimed to be present in these superconductors.
More experiments are also required to fully understand the origin and 
the physics behind such additional terms.

\newpage
\center {\Large \bf Figure captions}
\vspace{0.5cm} 
\begin{itemize} 

\item[Fig. 1.] Normalized (by $\Delta(0)$ at $\alpha=1$) gap versus 
temperature for four different values of $\alpha$. The inset shows variation 
of the transition temperature with $\alpha$.

\item[Fig. 2.] The order parameter as a function of angle (measured with 
respect to the line $(\pi,\pi)-(\pi,0)$ in the first BZ as in 
refs.\cite{mesot,nor1}) with increasing mixture of the higher harmonic.  

\item[Fig. 3] The Density of States in the superconducting state (dotted
line) with $\alpha=1$ and in the normal state (solid line) at $\delta=0.17$. 
The inset shows the difference in the DOS for $\alpha=1$ and $
\alpha=0.8$ in the superconducting state close to the node at very low
temperature (the gaps were 15 meV and 11.8 meV as in Fig. 1). 
 
\item[Fig. 4.] The frequency- and momentum-averaged $A({\bf k}, \omega)$   
(panel 1 \& 2) and the corresponding EDCs  (panel 3, see text) at various 
angles (measured from $Y-\bar{M}$ direction) on the FS (shown outside the panel
in the Y-quadrant). Figures (a)-(e) are for angles 45, 35, 23, 10 and 0 degree. 
The solid and dotted lines correspond to $\alpha=1.0$ and 0.9.     

\item[Fig. 5] The specific heat curves at three different levels of
mixing of the higher harmonic ((a), (b) and (c) correspond to $\delta=0.10,  
0.17$ and 0.28).  The figures clearly reveal the enhanced quasiparticle 
excitations  as $\alpha$ deviates from one. 
 
\item[Fig. 6.] The superfluid density is shown against temperature for three
different values of $\delta=0.10,\, 0.17 $ and 0.28 (Figs. (a)-(c)). 
The change in slope at low temperatures is clearly visible.  
 
\end{itemize} 


\newpage 
\begin{thebibliography}{999}
\bibitem{van} D. J. van Harlingen, Rev. Mod. Phys. {\bf 67} 515 (1995).
\bibitem{leggett} J. Annett, N. Goldenfeld and A. J. Leggett in {\it Physical 
Properties of High Tempearture Superconductors}, vol. 5, ed. D. M. Ginsberg, 
(World Scientific) (1996).  
\bibitem{nor0} M. R. Norman et al., Nature (London), {\bf 392} 134 (1998).  
\bibitem{mesot} J. Mesot et al., cond-mat/9812377 (revised version, 17 March 
1999).
\bibitem{scal} D. J. Scalapino, Phys. Reports. {\bf 250}, 329 (1995). 
\bibitem{cgt} B. Chattopadhya, D. M. Gaitonde and A. Taraphder,
Europhys. Lett., {\bf 34}, 705 (1996). 
\bibitem{nor1} R. Fehrenbacher and M. Norman, Phys. Rev. Lett., {\bf 74}, 3884  
(1995); M. R. Norman, M. Randeria, H. Ding, J. C. Campuzano, Phys. Rev. B 
{\bf 52}, 615 (1995). 
\bibitem{moat} P. K. Mohanty and A. Taraphder, Cond-mat/9603180; J. Phys.
Condens. Matter, {\bf 10}, 621 (1998).  
\bibitem{kot} G. Kotliar, Phys. Rev. B {\bf 37} 3664 (1988).  
\bibitem{ding} H. Ding et. al., Phy. Rev. Lett. {\bf 74}, 2784(1995); M. R. 
Norman et al., Nature (London), {\bf 392},  157 (1998). 
\bibitem{scal1} D. J. Scalapino, S. R. White and S. C. Zhang, Phys. Rev. 
Lett., {\bf 68}, 2830 (1992). 
\bibitem{tobe} Tulika Maitra and A. Taraphder, {\it to be published}. 
\bibitem{moler} K. A. Moler et al., Phys. Rev. B {\bf 55}, 3954 (1997). 
\bibitem{ran} M. Randeria et al., Phys. Rev. Lett {\bf 74},  4951 (1995).   
\bibitem{hirs} P. J. Hirschfeld and N. Goldenfeld, Phys. Rev. B {\bf 48}, 
4219 (1993). 
\bibitem{hardy} W. N. Hardy, Phys. Rev. Lett {\bf  70}, 3999 (1993);
\bibitem{ren} Y. Ren, Phys. Rev. Lett, {\bf 74} 3680 (1995).
\bibitem{hirs1} P. J. Hirschfeld, Peter Wolfle and D. Einzel, Phys. Rev. B,
{\bf 37}, 83 (1988). 
\bibitem{cmv} C. M. Varma, preprint.
\bibitem{haas} Stefan Haas et al. Phys. Rev. B {\bf 56}, 5108 (1997); 
G. Haran and A. D. S. Nagi, Phys. Rev. B {\bf 58}, 12441 (1998).
 
\end{thebibliography}
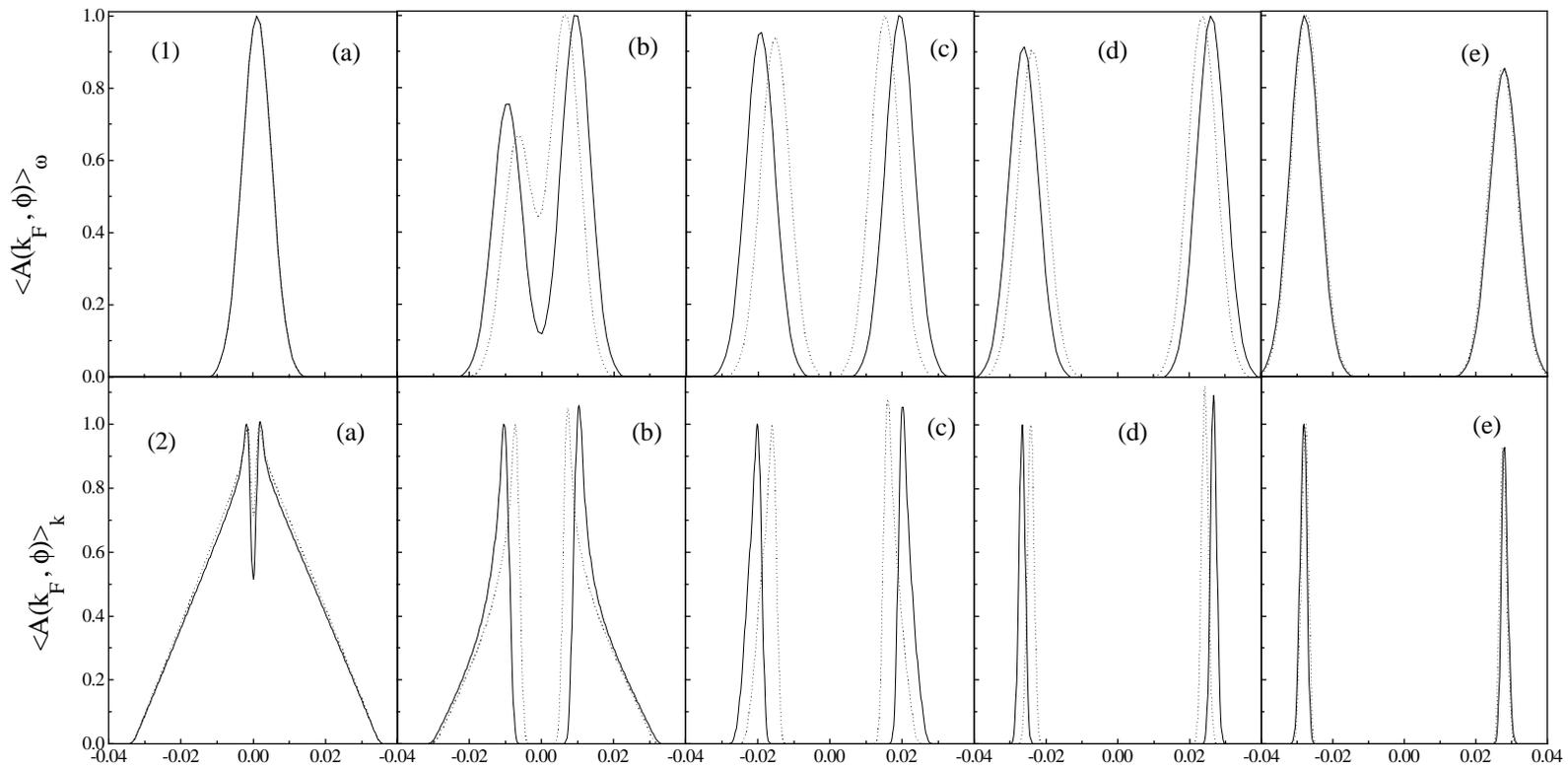
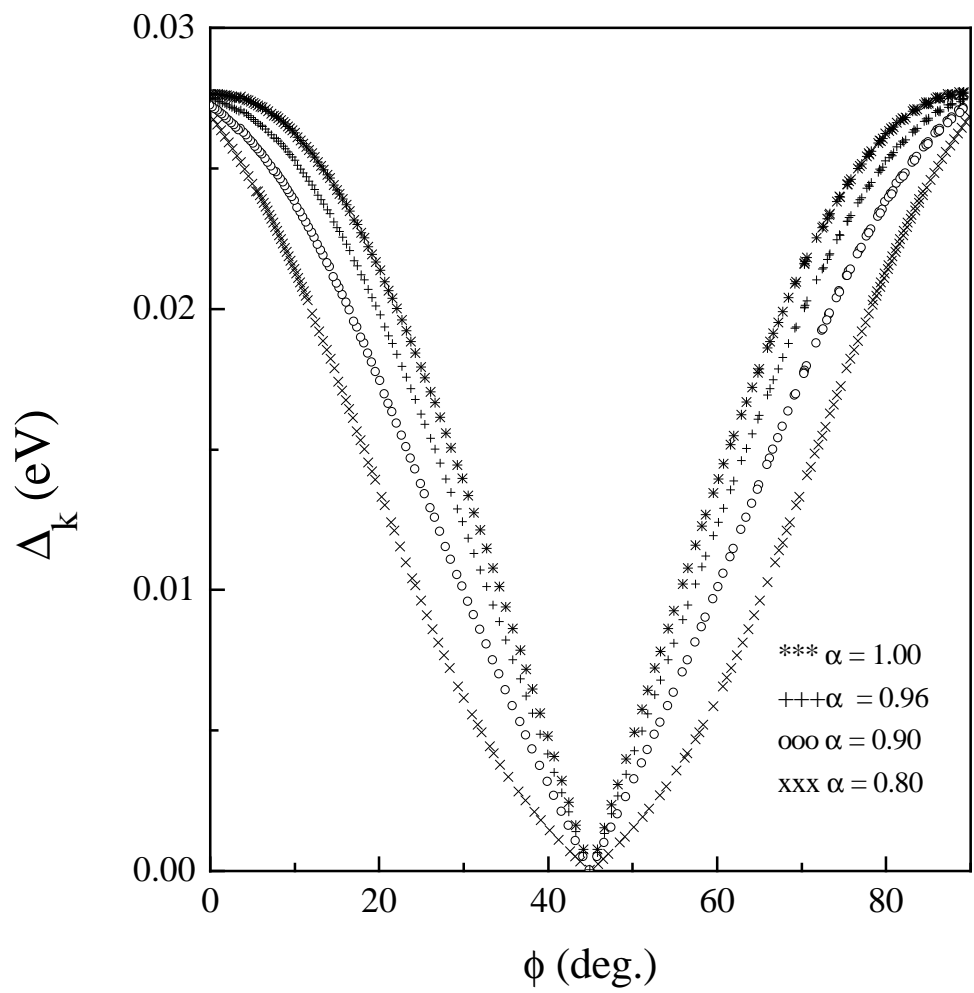
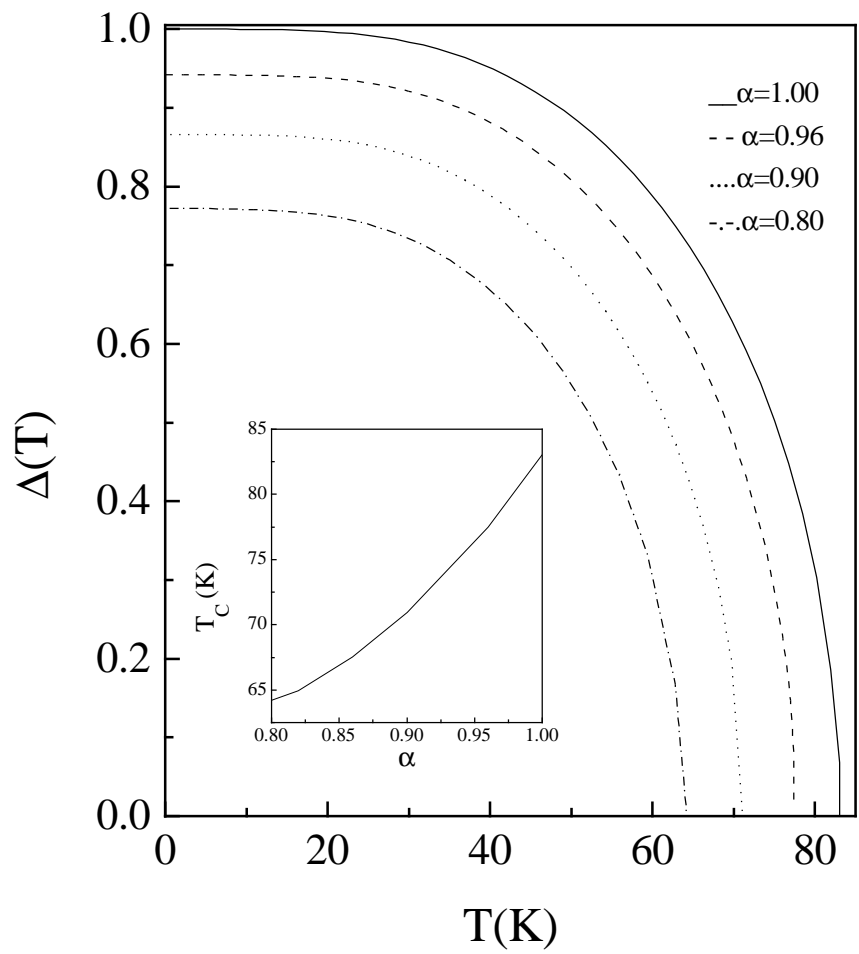
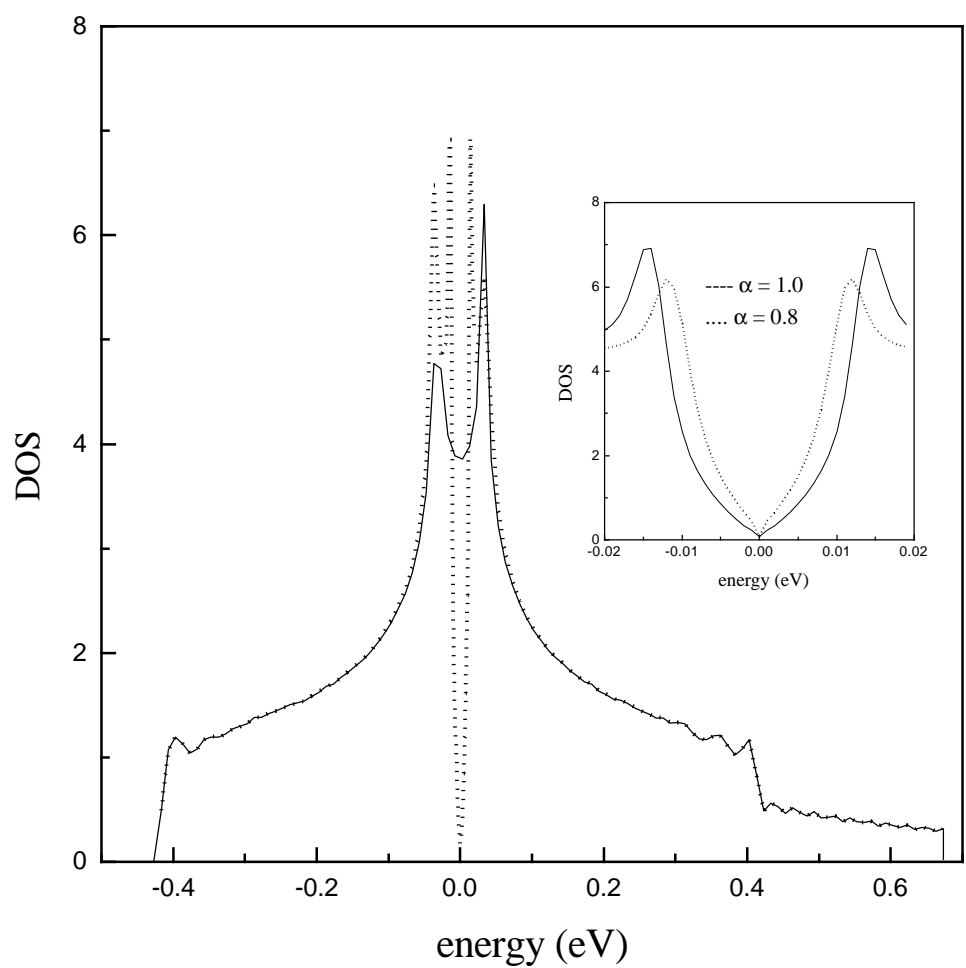
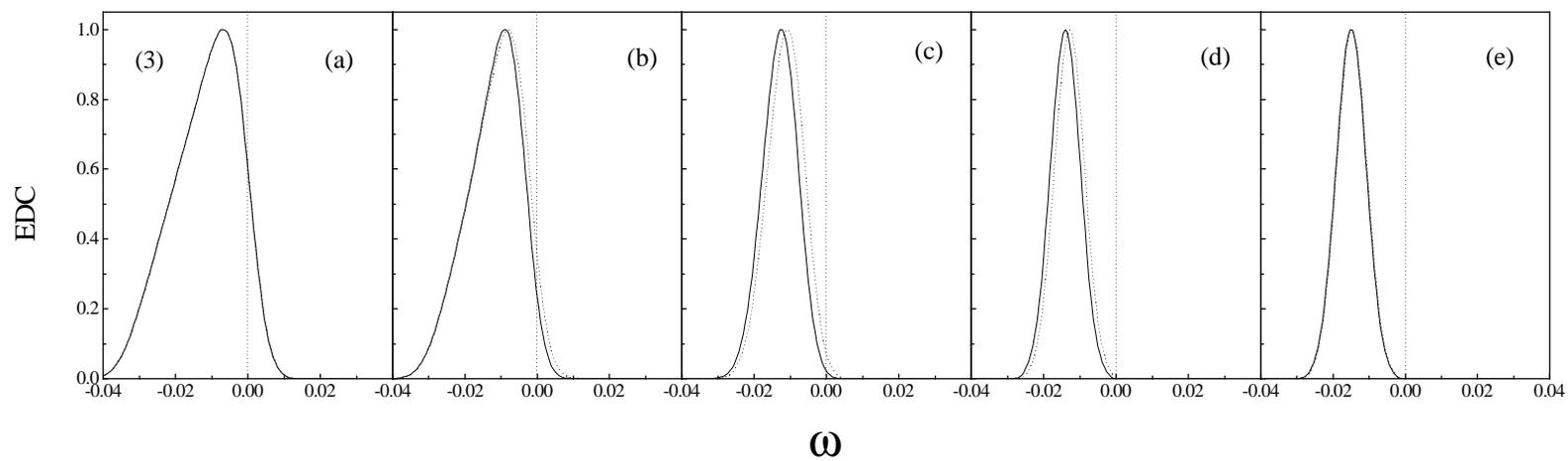
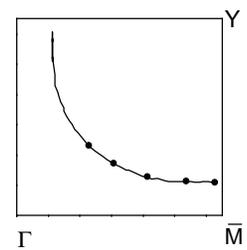
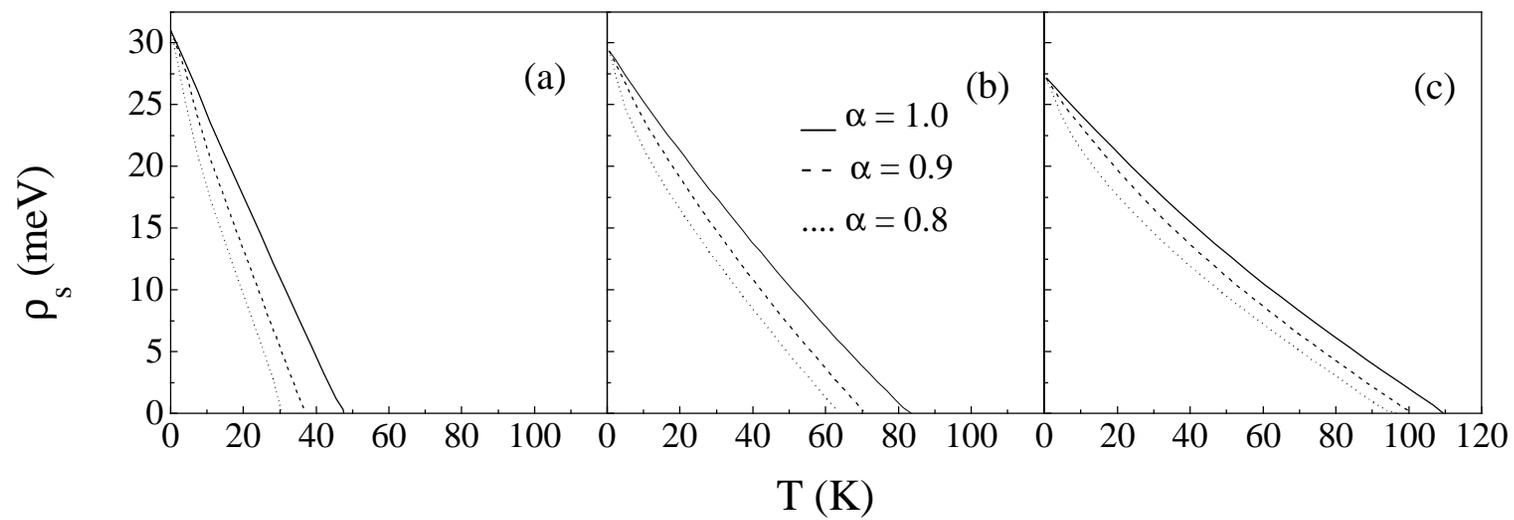
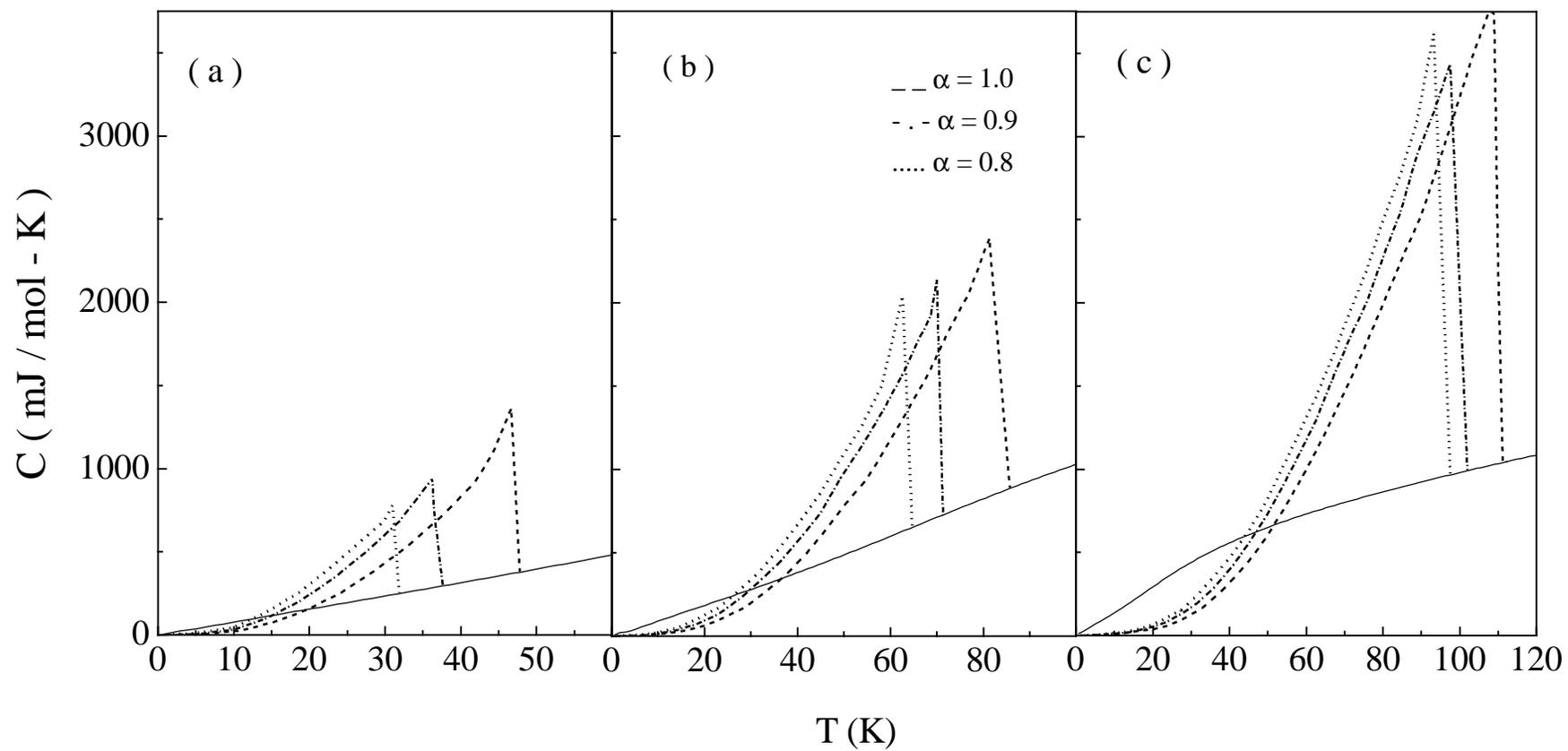
\end{document}